# Automatic Detection of COVID-19 and Pneumonia from Chest X- Ray using Deep Learning

Sarath Pathari[1]

**Abstract:**

In this study, a dataset of X-ray images from patients with common viral pneumonia, bacterial pneumonia, confirmed Covid-19 disease was utilized for the automatic detection of the Coronavirus disease. The point of the investigation is to assess the exhibition of cutting edge convolutional neural system structures proposed over the ongoing years for clinical picture order. In particular, the system called Transfer Learning was received. With transfer learning, the location of different variations from the norm in little clinical picture datasets is a reachable objective, regularly yielding amazing outcomes. The datasets used in this trial. Firstly, a collection of 24000 X-ray images includes 6000 images for confirmed Covid-19 disease,6000 confirmed common bacterial pneumonia and 6000 images of normal conditions. The information was gathered and expanded from the accessible X-Ray pictures on open clinical stores. The outcomes recommend that Deep Learning with X-Ray imaging may separate noteworthy biological markers identified with the Covid-19 sickness, while the best precision, affectability, and particularity acquired is 97.83%, 96.81%, and 98.56% individually.

**Keywords** COVID19· Bacterial Pneumonia Detection· Viral Pneumonia Detection· Deep learning· Transfer learning

## 1. Introduction

Since December 2019, a novel coronavirus (SARS-CoV2) has spread from Wuhan to the whole of China, and many other countries. By April 18, more than 2 million confirmed cases, and more than 150,000 deaths cases were reported in the world [1]. Because of the inaccessibility of helpful treatment or immunization for novel COVID-19 sickness, the early conclusion is of genuine significance to give the chance of prompt separation of the speculated individual and to diminish the opportunity of disease to sound populace.

Sarath Pathari

Turn around translation polymerase chain response (RT-PCR) or quality sequencing for respiratory or blood examples are presented as principle screening techniques for COVID-19. Be that as it may, the all-out constructive pace of RT-PCR for throat swab tests is accounted for to be 30 to 60 percent, which as needs be respected un-analyzed patients, which may infectiously taint a tremendous populace of sound individuals. Chest radiography imaging (e.g., X-beam or figured tomography (CT) imaging) as a standard apparatus for pneumonia conclusion is anything but difficult to perform with quick analysis. Chest CT has a high affectability for the finding of COVID-19 [2] and X-beam pictures show visual lists associated with COVID-19. The reports of chest imaging showed multilobar inclusion and fringe airspace opacities.

The opacities most now and again announced are ground-glass (57%) and blended lessening (29%) [3]. During the early course of COVID-19, ground glass design is found in regions that edge the pneumonic vessels and might be hard to acknowledge outwardly. Lopsided sketchy or diffuse airspace opacities are additionally announced for COVID-19 [4]. Such unobtrusive variations from the norm must be deciphered by master radiologists.

Thinking about the enormous pace of associated individuals and predetermined number with prepared radiologists, programmed strategies for the ID of such unobtrusive variations from the norm can help the conclusion system and increment the pace of early analysis with high exactness. Man-made consciousness (AI)/AI arrangements are conceivably incredible assets for taking care of such issues.

## 2. Impact Statement

In this paper, We identify whether a person had COVID-19, Bacterial Pneumonia, Viral Pneumonia and Healthy lungs or not. We Identify this by using Transfer Learning which is the most widely used method and hence we use the same to get a better accuracy. This application would be very useful for hospitals for the identification COVID-19 in a cheap and efficient way as it would save a lot of money and deployment cost. It would also help the people and the society.

## 3. Dataset

We have utilized the x-ray pictures from 4 datasets, to make the Chest X-ray dataset. The Chest X-ray dataset contains 24096 preparing train images and 3,615 test images. One of the utilized datasets is the as of late distributed Mendeley Chest X-ray Dataset[6], which contains a lot of pictures from distributions on Normal, Bacterial Pneumonia, and Viral Pneumonia themes, gathered by Daniel Kermany. This dataset contains a blend of chest Xray and CT pictures. We also further augmented it to achieve enhanced accuracy.

We also used the ieee8023 dataset [5] which contains 254 COVID-19 images in different views(PA, AP, L, AP-SUPINE, AP-SEMI ERECT). Since we had a very less number of COVID-19 images we had to augment it up to 6000 images.

It is referenced that this dataset is consistently refreshed. It additionally contains some meta-information about every patient, for example, sex and age. Our COVID-19 pictures are largely originating from this dataset. Consequently, we have picked 900 (This included augmented images as well) COVID-19 pictures to remember for the test set and 5100 COVID-19 pictures for the preparation set.

Information increase is applied to the preparation set to expand the quantity of COVID19, Normal, Bacterial Pneumonia, Viral Pneumonia tests to 6000(by a mix of flipping, pivot, little twisting, and over-examining). We ensured all pictures for every patient go just to one of the preparations or test sets.

For the non-COVID(Normal) tests in the preparation set, we just utilized pictures having a place to a solitary sub-classification, made out of 6096 pictures which also includes augmented images. For Pneumonia, we classified into 2 sub-labels: Viral and Bacterial Pneumonia respectively. We gathered around 8546 and we augmented those images to make it up to 12000.

**Table 1** The specific number of pictures of each class for both preparing and testing.

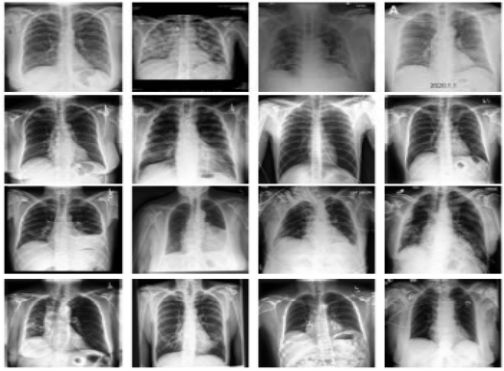

**Fig. 1.** Images from the COVID X-Ray dataset.

## 4. The Proposed Framework

Since so far, the number of publicly available images, which are labeled as COVID-19 is very limited, it may not be possible to train a deep CNN from scratch to detect COVID-19 from X-ray images. To overcome this issue, we use a well-known strategy in machine learning, called "transfer learning", and fine-tune four popular pre-trained deep neural networks on the training images of COVID-19 X-Ray Dataset. We will first provide a quick introduction of transfer learning and then discuss the proposed framework.

### 4.1 Transfer Learning Approach

In transfer learning, a model trained on one task is re-purposed on another related task, usually by some adaptation toward the new task. For example, one can imagine using an image classification model trained on MobileNet (which contains millions of labeled images) to initiate task-specific learning for COVID-19 detection on a smaller dataset. Transfer learning is mainly useful for tasks where enough training samples are not available to train a model from the beginning, such as medical image classification for rare or emerging diseases, in which adequately huge quantities of marked examples may not be accessible. This is particularly the situation for

| DATASET | NORMAL | COVID-19 | BACTERIAL PNEUMONIA | VIRAL PNEUMONIA |
|---|---|---|---|---|
| TRAINING SET | 4096(After Augmentation:6096) | 254(After Augmentation:6000) | 5560(After Augmentation:6000) | 2986(After Augmentation:6000) |
| TEST SET | 915 | 900 | 900 | 900 |

models dependent on profound neural systems, which have countless parameters to prepare. By utilizing move learning, the model parameters start with a; prepared decent introductory qualities that lone need some little adjustments to be better curated toward the new undertaking.

There are two primary manners by which the pre-prepared model is utilized for an alternate undertaking. In one methodology, the pre-prepared model is treated as an element extractor (i.e., the interior loads of the pre-prepared model are not adjusted to the new errand), and a classifier is prepared on it to perform arrangement. In another methodology, the entire system, or a subset thereof, is calibrated on the new errand. Along these lines, the pre-prepared model loads are treated as the underlying qualities for the new errand and are refreshed during the preparation stage.

In our case, since the number of images in the COVID-19 category is very limited, we only fine-tune the last layer of the convolutional neural networks, and essentially use the pre-trained models as a feature extractor. We assess the presence of a pre-prepared model which is MobileNetV3 [7]. In the following segment, we give a fast review of the design of the MobileNet model, and how they are utilized for COVID-19 acknowledgment.

### 4.2 COVID-19 Detection Using MobileNetV3

The bleeding edge CNN called Mobile Net For the requested task, the top tier CNN called Mobile Net was used. In their work, the makers indicated the predominance of Mobile Net in diminishing the False Negatives for the revelation of COVID-19, diverged from other 7 eminent CNNs. Furthermore, this CNN presents a less number of parameters that stood out from various CNNs, which makes it appropriate for brisk getting ready.

The Mobile Net [8] model relies upon depth-wise detachable convolutions, which is a kind of convolutions changing a customary convolution into a depth wise convolution and a 1x1 convolution, which is normally known as pointwise convolution. This strategy reduces the number of parameters unquestionably. To the most noteworthy purpose of the Mobile Net v2, a Global Average Pooling layer was incorporated, which unquestionably lessens the issue of overfitting.

The isolated picture features are inserted into a Neural Network of 2500 center points to perceive the unnecessary and the basic ones. To also help to the overfitting decline, the heaps of every component are normalized utilizing a Batch Normalization layer, while we self-rulingly zero out the half of the yields of neurons at sporadic, through a Dropout layer. The point of the particular assessment isn't simply to achieve a high request accuracy yet to achieve this by means of setting up the CNN without any planning. This strategy is alluring over trade making sense of how to survey the significance of the features removed from the specific pictures, while not depending upon features recently learned by the pre-arranged model, the fundamental getting ready of which was performed utilizing non-clinical pictures. Considering the results, the removed features may be evaluated to assume that they may set up veritable biological markers related to various afflictions.

Fig. 2. Review of the component extraction process.

## 5. Experimental Results

### 5.1 Model Hyper-Parameters

We adjusted each model for 2000 epochs. The gathering size is set to 16, and the ADAM enhancer is used to propel the setback work, with a learning pace of 0.001. All photos are down-tried to 224x224 before being dealt with to the neural framework (as these pre-arranged models are commonly arranged with a specific picture objective).

### 5.2 Evaluation Metrics

Metrics The measurements, in light of which the assessment of the presentation is made, are the general 4-class accuracy(Healthy versus COVID-19 versus Bacterial Pneumonia versus Viral Pneumonia). Since the present test dataset is profoundly imbalanced (as there are 120 pictures with COVID-19, and 4000 pictures that are Non-COVID and 1800 bacterial and 800 viral pneumonia pictures before augmentation), affectability and particularity are two proper measurements which can be utilized for revealing the model execution. These measurements are additionally generally utilized in clinical space.

**Sensitivity**= $\frac{\#Images\ correctly\ predicted\ images}{\#Total\ Actual\ Images}$

## 6. Results

In this segment, the outcomes for the diverse examination arrangements are introduced. In light of those outcomes, the ideal system is chosen, and presumptions are made concerning its adequacy.

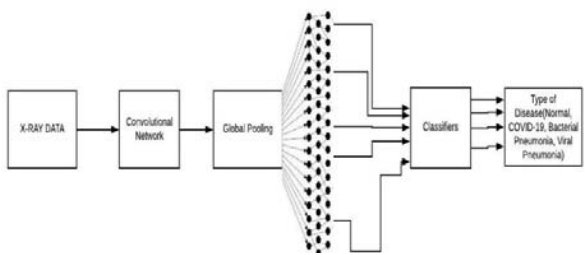

**Table 2.** Confusion Matrix for the 4-class classification employing transfer learning with of-the-self features.

**Table 3.** Accuracy, Sensitivity, and Specificity when the training from the scratch strategy was followed

| NETWORK | ACCURACY 4-CLASS(%) | SENSITIVITY(%) | SPECIFICITY(%) |
|---|---|---|---|
| Mobile Net v3 | 95.58 | 97.52 | 95.14 |

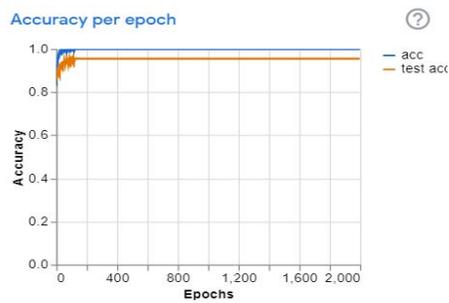

**Fig. 3.** Accuracy Per Epoch.

From the above graph, we can infer that we have achieved 99% Train Accuracy and 95.58 Test Accuracy.

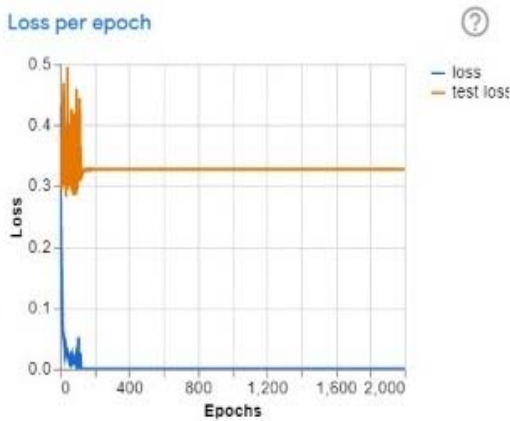

**Fig. 4.** Loss Per Epoch.

From the above graph, we can infer that there was a Train Loss of 0.0000125 and a Test Loss of 0.3245.

## 7. Conclusion

The commitment to this work is two-overlay. Right

| PREDICTED LABELS | ACTUAL LABELS | | | |
|---|---|---|---|---|
| | ACTUAL NORMAL | ACTUAL COVID-19 | ACTUAL BACTERIAL PNEUMONIA | ACTUAL VIRAL PNEUMONIA |
| PREDICTED NORMAL | 5760 | 20 | 41 | 32 |
| PREDICTED COVID-19 | 4 | 5940 | 19 | 5 |
| PREDICTED BACTERIAL PNEUMONIA | 124 | 30 | 5579 | 323 |
| PREDICTED VIRAL PNEUMONIA | 112 | 10 | 362 | 5640 |

off the bat, minimal effort, quick, and programmed discovery of the COVID-19 malady was accomplished, using a fundamentally huge example 13 of a few respiratory diseases. Moreover, the assessment prescribes that future research should be directed to investigate the possible lead of the removed feature as Bio-markers since there is sufficient evidence, considering the particular results. Besides, the benefit of programmed recognition of COVID-19 from either clinical picture lies in the decrease of presentation of nursing and clinical staff to the flare-up. This study is conducted on a set of publicly available images, which contains less than 200 COVID-19 images, and more than 3,000 non-COVID images. Due to the limited number of COVID-19 images publicly available so far, further experiments are needed on a larger set of cleanly labeled COVID-19 images for a more reliable estimation of the sensitivity rates.

**Compliance with ethical standards**

Conflict of interest: None.

## References

[1] http://www.worldometers.info

[2] Hongyan Hou, Zhenlu Yang, Ai, Tao, Chenao Zhan.
"Correlation of chest CT and RT-PCR testing in coronavirus

disease 2019 (COVID-19) in China: a report of 1014 cases."
Radiology (2020): 200642.

[3] Prachi P. Agarwal, Weifang, Kong, and. "Chest imaging appearance
of COVID-19 infection." Radiology: Cardiothoracic
Imaging 2, no. 1 (2020): e200028.

[4] Rodrigues, J.C. et al. An update on COVID-19 for the radiologist
- British Society of Thoracic Imaging statement. (2020)
Clinical Radiology.

[5] COVID-19 Image Data Collection
https://arxiv.org/abs/2003.11597

[6] https://data.mendeley.com/datasets/rscbjbr9sj/3

[7] Kaiming He, "Deep residual learning for image recognition."
Proceedings of the IEEE conference on computer vision
and pattern recognition. 2016.

[8], Matthew W.Iandola, Forrest N., Kurt Keutzer. "SqueezeNet:AlexNet-level accuracy with 50x fewer parameters and¡ 0.5 MB model size." arXiv preprint arXiv:1602.07360 (2016).

[9] http://codelabs.developers.google.com